\documentclass[12pt]{article}

\usepackage{amssymb}

\newcommand{\sca}{\ensuremath{\mathcal{A}}}
\newcommand{\scb}{\ensuremath{\mathcal{B}}}
\newcommand{\scd}{\ensuremath{\mathcal{D}}}

\newcommand{\sch}{\ensuremath{\mathcal{H}}}
\newcommand{\scs}{\ensuremath{\mathcal{S}}}
\newcommand{\scx}{\ensuremath{\mathcal{X}}}

\newcommand{\sone}{\ensuremath{\mathcal{S}_{1}}}

\newcommand{\dera}{\ensuremath{Der(\mathcal{A})}}

\newcommand{\phst}{\ensuremath{\Phi_*}}
\newcommand{\phstup}{\ensuremath{\Phi^*}}

\newcommand{\oa}{\ensuremath{\mathcal{O}(\mathcal{A})}}

\begin{document}
\begin{center} \textbf{\large A Stepwise Planned Approach to the Solution
of Hilbert's Sixth Problem. III : Measurements and von Neumann
Projection/Collapse Rule}

\vspace{.15in} \textbf{Tulsi Dass}

 Indian Statistical Institute, Delhi Centre, 7, SJS
Sansanwal Marg, New Delhi, 110016, India.

  E-mail: tulsi@isid.ac.in; tulsi@iitk.ac.in
\end{center}

\vspace{.18in}  Supmech, the universal mechanics developed in the previous two papers, accommodates both quantum and classical mechanics as subdisciplines (a brief outline is included for completeness); this feature facilitates, in a supmech based treatment of quantum measurements, an unambiguous treatment of the apparatus as a quantum system approximated well by a classical one. Taking explicitly into consideration the fact that observations on the apparatus are made when it has `settled down after the measurement interaction' and are restricted to macroscopically distinguishable pointer readings, the unwanted superpositions of (system + apparatus) states are shown to be suppressed; this provides a genuinely physics based justification
for the (traditionally \emph{postulated}) von Neumann projection/collapse rule. The decoherence mechanism brought into play by the stated observational constraints is free from the objections against the traditional decoherence program.

\vspace{.15in} \noindent PACS : 0365.Ta; 03.65.Ca; 03.65.Yz; 03.65.Sq

\newpage

\vspace{.2in} \noindent \textbf{\large 1. Introduction}

\vspace{.1in} In this open-ended program aimed at a solution of
Hilbert's sixth problem (relating to axiomatization of physics and
probability theory), the first two papers (Dass [1,2];
henceforth referred to as I and II) were devoted to evolving the
geometro-statistical framework of a universal mechanics called
`supmech' and a consistent autonomous treatment of quantum systems in that
framework. In this third paper, we shall treat measurements on
quantum systems in the supmech framework and obtain a
straightforward derivation of the von Neumann projection/collapse
rule, obtaining, in the process, a clear understanding of the sense
in which this rule should be understood.

The negative result about the possibility of a consistent
quantum-classical interaction in the supmech framework [obtained as
part of theorem (2) in I; see section 4.4 below] is by no means
`fatal' for a consistent treatment of measurement interaction
between the system and apparatus. It turns out that it is adequate
to treat the apparatus carefully as a quantum system
approximated well by a classical one (in the setting of, for example,
phase space descriptions of quantum and classical dynamics); the fact
that both quantum and classical mechanics are special subdisciplines
of supmech is very helpful in such a treatment. We shall see that,
taking properly into consideration (i) the `settling down of the
apparatus after the measurement interaction', and (ii) the fact that
the observations on the apparatus are restricted to macroscopically
distinguishable pointer readings (this is what \emph{automatically}
brings into play the decohering effect of the internal environment
of the apparatus), the unwanted  superpositions of (system +
apparatus)-states can be shown to be suppressed, leading eventually
to the projection/collapse rule postulated in von Neumann's
treatment  of measurements [3].

In the next section, the measurement problem in quantum mechanics
(QM) is recalled. In section 3, some proposed improvements in the
treatment of the physics of the apparatus are outlined. In section
4, we briefly recall the essential features of supmech and, in
section 5, a supmech-based treatment of measurements on a quantum
system is given leading eventually to the von Neumann projection
rule; the functioning of some crucial ingredients of this treatment
is illustrated with the example of the Stern-Gerlach experiment in
section 6. In section 7, the present work is compared with the
traditional decoherence program. In section 8, we add, to the
list of seven axioms of the
supmech program given in II, another one covering measurement
situations. The last section contains some concluding remarks.

\vspace{.15in} \noindent  \textbf{\large 2. The Measurement Problem in
Quantum Mechanics}

\vspace{.1in} We consider, for simplicity, the measurement of an
observable (of a quantum system S) represented by a self-adjoint
operator F (acting in an appropriate dense domain in the Hilbert space
$\sch_S$ of S) having a non-degenerate spectrum with the eigenvalue
equations $F |\psi_j> = \lambda_j |\psi_j>$ (j = 1,2,...). The
apparatus A is chosen such that, to each of the eigenvalues
$\lambda_j$ corresponds a pointer position $M_j$. If the system is
initially in an eigenstate $|\psi_j>$, the apparatus is supposedly
designed to give, after the measurement interaction, the pointer
reading $M_j$; the outcome of the measurement is then understood as
$\lambda_j$. A question immediately arises : `What is the
measurement outcome when the initial state of the system S is a
superposition state $|\psi> = \sum_j c_j |\psi_j> $ ?' The
theoretical framework employed for the treatment of measurements on
quantum systems must provide a satisfactory answer to this question.

 The standard treatment of measurements in QM (von Neumann [3];
Wheeler and Zurek [4]; Jauch [5]; Omnes [6]; Dass [7]) is
due to von Neumann who, treating the apparatus as a quantum system,
introduced, for the pointer positions $M_j$, state vectors $|\mu_j>$
in the Hilbert space $\sch_A$ of the apparatus. The Hilbert space
for the coupled system (S + A) is taken to be the tensor product
$\sch = \sch_S \otimes \sch_A$. The measurement interaction is
elegantly described (Omnes [6]; Dass [7]) by a unitary operator
U on \sch \ which, acting on the initial state of (S+A) (with the
system S in the initial state in which it is prepared for the
experiment and the apparatus in the `ready' state which we denote as
$|\mu_0>$) gives an appropriate final state. We shall assume the
measurement to be \emph{ideal} which is supposedly such that
(Omnes [8]) `when the measured system is initially in an eigenstate of the
measured observable, the measurement  leaves it in the same state.'
In this case, the measurement outcome must be the corresponding
eigenvalue which must be indicated by the final pointer position.
This implies
\begin{eqnarray}
U(|\psi_j> \otimes |\mu_0>) = |\psi_j> \otimes |\mu_j>.
\end{eqnarray}
For S in the initial state $ |\psi> = \sum c_j|\psi_j>$, the final
(S + A)- state must be, by linearity of U,
\begin{eqnarray}
|\Psi_f> \  \equiv U[(\sum_j c_j |\psi_j>) \otimes |\mu_0>] = \sum_j
c_j [|\psi_j> \otimes |\mu_j>].
\end{eqnarray}
Note that the right hand side of Eq.(2) is a superposition of the
quantum states of the (generally \emph{macroscopic}) system (S + A).

Experimentally, however, one does not observe such superpositions.
Instead, one obtains, in each measurement, a definite outcome
$\lambda_j$ corresponding to the final (S + A)-state $|\psi_j>
\otimes |\mu_j>$. Repetitions of the experiment, with system in the
same initial state, yield various outcomes randomly such that, when
the number of repetitions becomes large, the relative frequencies of
various outcomes tend to have fixed values. To account for this, von
Neumann postulated that, after the operation of the measurement
interaction as above, a discontinuous, noncausal and instantaneous
process takes place which changes the state $ |\Psi_f>$ to the state
represented by the density operator
\begin{eqnarray}
\rho_f & = &\sum_i \tilde{P}_i |\Psi_f><\Psi_f| \tilde{P}_i \\
       & = & \sum_j |c_j|^2 [|\psi_j><\psi_j| \otimes
       |\mu_j><\mu_j|];
\end{eqnarray}
here $\tilde{P}_i = |\psi_i><\psi_i| \otimes I_A$ where $I_A$ is the
identity operator on $\sch_A$. This is referred to as von Neumann's
\emph{projection postulate} and the phenomenon with the above
process as the underlying process the \emph{state vector reduction}
or \emph{wave function collapse}. Eq.(4) represents, in the von
Neumann scheme, the (S +A)-state on the completion of the
measurement. It represents an ensemble of (S + A)-systems in which a
fraction $p_j = |c_j|^2$ appears in the j th product state in the
summand. With the projection postulate incorporated, the von Neumann
formalism, therefore, predicts that, in a measurement with the
system S initially in  the superposition state as above,

\noindent (i) the measured values of the observable F are the random
numbers $\lambda_j$ with respective probabilities $|c_j|^2$;

\noindent (ii) when the measurement outcome is $\lambda_j$, the
final state of the system is $|\psi_j>$.

\noindent Both the predictions are in excellent agreement with
experiment.

The main problem with the treatment of a quantum measurement given
above is the ad-hoc nature of the projection postulate. Moreover,
having to invoke a discontinuous, acausal and instantaneous process
is an unpleasant feature of the formalism. The so-called measurement
problem in QM is essentially the problem of explaining the final
state (4) without introducing anything ad-hoc and/or physically
unappealing in the theoretical treatment. This means that one should
either give a convincing dynamical explanation of the reduction
process or else circumvent it; we shall do the former.

A critical account of various attempts to solve the measurement
problem and related detailed references may be found in the author's
article [7]; none of them can be claimed to have provided a
satisfactory solution. [Even the relatively more appealing
decoherence program (Zurek [9]) has problems ([10,11,7];  see
section 7).]

\vspace{.15in} \noindent \textbf{\large 3. Doing Justice to the Physics of
the Apparatus}

\vspace{.1in}  Von Neumann's treatment does not do adequate justice
to the physics of the apparatus and needs some improvements. We
propose to take into consideration the following points :

\vspace{.1in} \noindent (i) The apparatus A is a quantum mechanical
system admitting, to a very good approximation, a classical
description. Even when the number of the effective apparatus degrees
of freedom is not large (for example, in the Stern-Gerlach
experiment, treated in section 6, where the center of mass position
vector of a silver atom acts as the effective apparatus variable), a
classical description of the relevant variables is adequate. This
feature must be properly
incorporated in the theoretical treatment to obtain a satisfactory
description of measurements. [Items (iii)-(v) below
cannot be treated properly unless this feature is incorporated.]
This is the right approach to avoid problems relating to
`quantum-classical divide' in the treatment of measurements on
quantum systems.

\vspace{.1in} \noindent (ii) Introduction of vector states for the
pointer positions is neither  desirable (no operational meaning can
be assigned to a superposition of the pointer states $|\mu_j>$) nor
necessary :  a better procedure is to introduce density operators
for the pointer states and take into consideration the fact that the
Wigner functions corresponding to them are approximated well by
classical phase space density functions.

\vspace{.1in} \noindent (iii) The pointer states have a stability
property : After the measurement interaction is over, the apparatus,
left to itself, settles quickly into one of the pointer positions.
It is this process which should replace von Neumann's
`instantaneous, non-causal and discontinuous' process.

\vspace{.1in} \noindent \emph{Note.} A detailed mathematical
treatment of this process, as we shall see below, is not necessary;
it is adequate to take its effect correctly into account. (The von
Neumann  projection postulate does this, but that is not enough;
one must give a physics based justification for the prescription.)
To get a feel for this, note that, in, for example, the Stern-Gerlach
experiment, treated in section VI, the measurement interaction is
over (ignoring fringe effects) after the atom is out of the region
between the magnetic pole pieces. In this case, by `the apparatus
settling to a pointer position' one means the movement of the atom
from just outside the pole pieces to a detector. In this case, the
choice of the detector is decided by the location of the atom just
after the measurement interaction. Details of motion of the atom
from the magnets to the detector is of no practical interest in the
present context. In the case of a macroscopic apparatus, the
`settling ...' refers to the process of the apparatus reattaining
(thermal, mechanical) equilibrium (disturbed slightly during the measurement
interaction) after the measurement interaction; again, details of
this process are not important in the present context. Different
pointer positions supposedly have [see axiom A8(d) in section 8]
mutually disjoint stability domains in
the phase space of the apparatus. The eventual
pointer position indicated in an experiment is decided by the
stability domain  in which the
point representing the classical state of the apparatus happens to
be immediately after the measurement interaction.

\vspace{.1in} \noindent (iv) Observations relating to the apparatus
are restricted to the pointer positions $M_j$. A properly formulated
dynamics (classical or quantum) which takes this into consideration
(treating the apparatus `respectfully' as a \emph{system}) would
involve, at appropriate stage, averaging over the inoperative part
of the phase space of the apparatus. It is this averaging, as we shall
see below, which [combined with item (v) below] produces the needful
decoherence effects  to wipe out undesirable quantum interferences.

\vspace{.1in} \noindent (v) Different pointer positions are
macroscopically distinguishable. We shall take this into
consideration by employing an appropriate energy-time uncertainty
inequality.

\vspace{.15in} \noindent \textbf{\large 4. Supmech : A Brief Outline}

\vspace{.1in} Supmech is an `all-embracing'  mechanics having both
classical and quantum mechanics as its subdisciplines. Its framework
facilitates an autonomous development of QM (i.e. without having to
quantize classical dynamical systems) and a transparent treatment of
quantum-classical correspondence. A brief presentations of its basic
features follows. Since fermionic objects are not needed in the
present work, we shall present only the bosonic version of supmech.

\vspace{.1in} \noindent \textbf{4.1 Probabilistic framework}

\vspace{.1in} \noindent \emph{a. Experimentally accessible systems.}
By these, we mean systems whose `identical' (for all practical
purposes) copies are reasonably freely available for repeated trials
of an experiment. Henceforth by a system we shall mean an
experimentally accessible one. Some aspects of systems not included
in this class (the universe and its `large' subsystems) can be
covered by the formalism of this paper with the slightly more
refined presentation of the basic postulates as given in  II and an
appropriate interpretation of classical probabilities in the
statistical analysis of the experimental data relating to such
systems.

\vspace{.1in} \noindent \emph{b. System algebra; Observables.}
Supmech associates, with every system S, a complex associative
topological $\star$-algebra $\sca = \sca^{(S)}$ having a unit
element (denoted here as I). [The topology is assumed to be locally convex
with some additional features (described in I); we shall skip the details.]
Observables of S are elements of the subset \oa \ of
Hermitian elements of \sca. A positive observable is a sum of terms
of the form $\sum_i A_i^*A_i$ where $A_i \in \sca$.

\vspace{.1in} \noindent \emph{c. States.} States of the system, also
referred to as the states of the system algebra \sca \ (denoted by
the letters $\phi, \phi^{\prime},..$), are defined as  continuous positive
linear functionals on \sca \ which are normalized [i.e. $\phi(I)
=1$]. The set of states of \sca \ will be denoted as \scs(\sca) and
the subset of pure states (i.e. those not expressible as nontrivial
convex combinations of other states) by $\sone (\sca)$. For any $A \in
\mathcal{O}(\sca)$ and $ \phi \in \scs (\sca)$, the quantity
$\phi(A)$ is to be interpreted as the expectation value of A when
the system is in the state $\phi$.

\vspace{.1in} \noindent \emph{d. Compatible completeness of
observables and pure states.} The pair (\oa, \sone(\sca)) is assumed
to be \emph{compatibly complete} in the sense that

\vspace{.1in} \noindent (i) given $A,B \in \oa, A \neq B, $ there
should be a state $\phi \in \sone(\sca)$ such that $ \phi(A) \neq
\phi(B)$;

\vspace{.1in} \noindent (ii) given two different states $\phi_1$ and
$\phi_2$ in \sone(\sca), there should be an $A \in \oa$ \ such that
$ \phi_1(A) \neq \phi_2(A)$.

\vspace{.1in} \noindent We shall refer to this condition as the `CC
condition' for the pair $ (\oa, \sone(\sca))$.

\vspace{.1in} \noindent \emph{e. Experimental situations and
probabilities.} In supmech, experimental situations relating to a
system with system algebra \sca \ are formalized in terms of
\emph{positive observable valued measures} (PObVMs) defined as
follows. One introduces a measurable space $(\Omega, \mathcal{F})$
where $\Omega$ is the `value space' (spectral space)  of one or more
observables and elements of $\mathcal{F}$ (measurable subsets) are
(standardized idealizations of) those subsets of $\Omega$ which can
be experimentally distinguished. A PObVM for the system S, based on
this measurable space, is a family $\{ \nu (E); E \in \mathcal{F}
\}$ where the objects $\nu (E)$  (\emph{supmech events}) are
positive observables in \sca \ such that
\[\begin{array}{l}
(i) \ \nu(\emptyset)=0, \ (ii) \  \nu(\Omega) = I,   \nonumber \\
(iii)\ \nu(\cup_i E_i) = \sum_i \nu(E_i)\  \textnormal{(for disjoint
unions)}.\end{array} \] It is the abstract counterpart of the
`positive operator-valued measure' (POVM) employed in Hilbert space
QM.  Given a state $\phi$ of the system S, we have a probability
measure $p_{\phi}$ on $(\Omega, \mathcal{F})$ given by
\begin{eqnarray}
p_{\phi}(E) = \phi(\nu(E)) \ \ \forall E \in \mathcal{F}.
\end{eqnarray}
Eq.(5) represents the theoretically desirable relationship between
supmech expectation values and classical probabilities. In supmech,
all probabilities in the formalism (which relate to statistics of
measurement outcomes) are stipulated to be of this type
(i.e. expectation values of supmech events).

\vspace{.12in} \noindent \textbf{4.2 Noncommutative symplectic
geometry}

\vspace{.1in} \noindent \emph{a. Derivation based noncommutative
differential calculus} (Dubois-Violette [12,13]).  Replacing, in
the classical differential form calculus, the commutative algebra
$C^{\infty}(M)$ of smooth complex valued functions on a manifold M
by an algebra \sca \ in the class mentioned above and the Lie
algebra $\scx (M)$ of smooth complex vector fields on M by the Lie algebra
\dera \ of derivations of \sca, the elements of the space
$\Omega^p(\sca)$ of noncommutative differential p-forms on \sca \
(p= 1,2,..) are defined as multilinear maps $(\dera)^p \rightarrow
\sca$ such that, for $\omega \in \Omega^p(\sca), \ X,Y \in \dera$
and $K \in Z(\sca)$, the center of \sca, we have
\begin{eqnarray*} \omega (..,X,Y,..) = - \ \omega (..,Y,X,..); \ \ \omega
( ..,K X,..) = K \omega (..,X,..); \end{eqnarray*} moreover,
$\Omega^0 (\sca) = \sca$. The usual operations on differential forms
[exterior product $\wedge : \Omega^p(\sca) \times \Omega^q (\sca)
\rightarrow \Omega^{p+q}(\sca)$, exterior derivative $d : \Omega^p
(\sca) \rightarrow \Omega^{p+1} (\sca)$, interior product $i_X :
\Omega^p (\sca) \rightarrow \Omega^{p-1} (\sca)$ and Lie derivative
$L_X : \Omega^p (\sca) \rightarrow \Omega^p (\sca)$ are defined
along lines parallel to the commutative case and analogous relations
involving these operations hold, with very few exceptions. [The not
so well known \emph{algebraic} definition of the exterior derivative
in the commutative case (Matsushima [14]) works as such for the
noncommutative case.] A differential form $\alpha$ is said to be closed
if $d \alpha = 0$ and exact if $\alpha = d \beta$ for some form $\beta.$

\vspace{.1in} \noindent \emph{b. Induced mappings on derivations and
differential forms.} These are analogues of the push-forward and
pull-back mappings on vector fields and differential forms induced
by diffeomorphisms between manifolds. Given a topological $\star$-algebra
isomorphism $\Phi : \sca \rightarrow \scb$, we have the induced linear mappings
$\phst : \dera \rightarrow Der (\scb)$ and $\phstup : \Omega^p
(\scb) \rightarrow \Omega^p (\sca)$ given by
\begin{eqnarray*} (\phst X)(B) & = & \Phi(X[\Phi^{-1}(B)]) \
\textnormal{for all} \ X \in \dera \ \textnormal{and} \ B \in \scb;
\\ (\phstup \omega)(X_1,..,X_p) & = & \Phi^{-1} [ \omega ( \phst X_1,..,
\phst X_p)] \\ & \ & \textnormal{for all} \  \omega \in \Omega^p
(\scb)\ \textnormal{and} \ X_1,..,X_p \in \dera. \end{eqnarray*}
These mappings satisfy the relations [with $\Psi : \scb \rightarrow
\mathcal{C}$ and other obvious notation] \begin{eqnarray*}  \ (\Psi
\circ \Phi)_* & = & \Psi_* \circ \Phi_*; \ \
 \ \Phi_* [X,Y] = [\phst X, \phst Y]; \  (\Psi \circ \Phi)^* = \Phi^*
\circ \Psi^*; \\ \phstup (\alpha \wedge \beta) & = & (\phstup
\alpha) \wedge (\phstup \beta); \ \phstup (d \alpha ) = d ( \phstup
\alpha ). \end{eqnarray*}

\vspace{.1in} \noindent \emph{c. Symplectic structures; Poisson
brackets} (Dubois-Violette [12,13]); \emph{Canonical
transformations.} The system algebra \sca \ is assumed to be
equipped with a symplectic form $\omega$ which, by definition, is a
closed 2-form which is non-degenerate in the sense that, for any $A
\in \sca$, there is a unique derivation $Y_A \in \dera$ such that
\[ i_{Y_A} \omega = - d A .\]
The pair $(\sca, \omega)$ is called a \emph{symplectic algebra}. For
any two elements A,B of \sca, their Poisson bracket (PB) is defined
as $\{ A, B \} = Y_A (B)$ and has the usual properties of
bilinearity, antisymmetry, Leibnitz rule and Jacobi identity.

Given two symplectic algebras $(\sca, \omega)$ and $(\sca^{\prime},
\omega^{\prime})$, a topological $\star$-algebra isomorphism
$\Phi : \sca \rightarrow
\sca^{\prime}$ is called a \emph{symplectic mapping} if $\phstup
\omega^{\prime} = \omega$. A symplectic mapping of $(\sca, \omega)$
onto itself is called a canonical/symplectic transformation. An
infinitesimal transformation of \sca \ of the form
\begin{eqnarray} A \mapsto A + \delta A; \ \delta A = \epsilon \{G,
A \} \end{eqnarray} is a canonical transformation (generated by $G
\in \sca$).

\vspace{.1in} \noindent \emph{d. Special algebras; the canonical
symplectic form.} An algebra \sca \ of the above mentioned type is
called special if all its derivations are inner (i.e. those of the
form $D_A$ with $D_A(B) = [A,B]$). The differential 2-form
$\omega_c$ defined on such an algebra \sca \ by
\begin{eqnarray}
\omega_c (D_A, D_B) = [A,B]
\end{eqnarray}
is said to be the \emph{canonical form} on \sca. It is a symplectic
form giving, for $A,B \in \sca$, $Y_A = D_A$ and $\{A, B \} = [A,
B].$ If one takes, on such an algebra, the form $\omega = b \
\omega_c$ as a symplectic form (where b is a nonzero complex
number), we have
\begin{eqnarray}
Y_A = b^{-1} D_A ,  \hspace{.4in} \{A, B \} = b^{-1}[A,B].
\end{eqnarray}
The \emph{ quantum Poisson bracket}
 \begin{eqnarray} \{ A, B \}_Q = (-i \hbar)^{-1} [A,B]
 \end{eqnarray}
is a special case of this with $b = -i \hbar;$ the corresponding
symplectic form is the \emph{quantum symplectic form} $\omega_Q = -
i \hbar \omega_c.$

\vspace{.12in} \noindent \textbf{4.3 Dynamics.}

\vspace{.1in} Dynamics in supmech is described (in the Heisenberg
type picture) by a one-parameter family $\Phi_t$ of canonical
transformations generated by by an observable $H \in \oa$ called the
Hamiltonian. Writing $\Phi_t (A) = A(t)$ and taking G = H and
$\epsilon = \delta t$ in Eq.(6),  we have the \emph{supmech
Hamilton's equation}
\begin{eqnarray} \frac{d A(t)}{dt} = \{ H, A(t) \} \equiv
\partial_H (A(t)). \end{eqnarray}
In the Schr$\ddot{o}$dinger type picture, time evolution is carried
by states, the two descriptions being related as
\begin{eqnarray} <\phi, A(t)> = <\phi(t), A> \end{eqnarray}
[so that $\phi(t) = \tilde{\Phi}_t (\phi)$ where the tilde indicates
transpose]. Writing $\phi(t + \delta t) = \phi(t) + \delta \phi(t)$,
we have, from equations (10,11), the \emph{supmech Liouville
equation} for the time evolution of states :
\begin{eqnarray} \frac{d \phi(t)}{dt} (A)= \phi(t) (\{ H, A \})
\equiv (\tilde{\partial}_H (\phi(t))(A). \end{eqnarray} We may
write, formally,
\begin{eqnarray} \Phi_t = exp(t \partial_H); \ \ \tilde{\Phi}_t = exp (t
\tilde{\partial}_H). \end{eqnarray}

The quadruple $\Sigma = (\sca, \sone(\sca), \omega, H)$ is called a
\emph{supmech Hamiltonian system}. Another  supmech hamiltonian
system $\Sigma^{\prime}$  is said to be equivalent to $\Sigma$ if
there is a symplectic mapping $\Phi : (\sca, \omega) \rightarrow
(\sca^{\prime}, \omega^{\prime})$ such that $\Phi(H) = H^{\prime}$.
The states are then related through $\tilde{\Phi}$. When states are
not being considered, we may refer to a triple $(\sca, \omega, H)$
as a supmech Hamiltonian system.

\vspace{.12in} \noindent \textbf{4.4 Interaction between two systems
in supmech}

\vspace{.1in} Given two sytems $S_1$ and $S_2$ considered as supmech
Hamiltonian systems $\Sigma_i = (\sca^{(i)}, \sone(\sca^{(i)}),
\omega^{(i)}, H^{(i)})$  [the PBs in the two algebras will be
denoted as $\{ .,. \}_i$ (i=1,2)], we treat the coupled system $(S_1
+ S_2)$ as a supmech Hamiltonian system $\Sigma = (\sca,
\sone(\sca), \omega, H)$ where $\sca = \sca^{(1)} \otimes
\sca^{(2)}$,
\begin{eqnarray}
\omega & = & \omega^{(1)} \otimes I_2 + I_1 \otimes \omega^{(2)}, \\
H & = & H^{(1)} \otimes I_2 + I_1 \otimes H^{(2)} + H_{int}
\end{eqnarray}
where $I_i$ is the unit element of $\sca^{(i)}$ (i= 1,2) and,
typically,
\begin{eqnarray*}
H_{int} = \sum_{i=1}^{n} F_i \otimes G_i.
\end{eqnarray*}

The 2-form $\omega$ of (14) is closed but, according to theorem (2)
in I, is non-degenerate if and only if either both the algebras
$\sca^{(1)}$ and $\sca^{(2)}$ are commutative or both noncommutative
with their respective PBs proportional to commutators with the
\emph{same} proportionality constant i.e.
\begin{eqnarray}
\{ A, B \}_1 = i \lambda [A, B]; \ \ \{ C, D \}_2 = i \lambda [C, D]
\end{eqnarray} (with the parameter $\lambda$ nonzero and real;
it can be chosen, by replacing the initially chosen symplectic
form by its negative, if necessary, to be positive). We make the
identification $\lambda = \hbar^{-1}$; the formalism, therefore,
\emph{dictates} the presence of a universal Planck type constant.

In both the
cases, the PB in \sca \ can be expressed in the form [I, Eq.(99)]
\begin{equation}
\{ A \otimes B , C \otimes D \}  =    \{ A,C \}_1 \otimes \frac{BD +
DB}{2} + \frac{AC + CA}{2} \otimes  \{ B,D \}_2.
 \end{equation}

The dynamics of the coupled system is governed, in the
Heisenberg type picture, by the supmech Hamilton's
equation [I, Eq.(101)]
\begin{eqnarray}
\frac{d}{dt} [ A(t) \otimes B(t)] & = & \{ H, A(t)
\otimes B(t)\} \nonumber \\
   & = & \{H^{(1)},A(t)\}_1 \otimes B(t)
                     + A(t) \otimes \{H^{(2)},B(t) \}_2
                     \nonumber \\
                    & \ &  + \{H_{int}, A(t) \otimes B(t) \}.
\end{eqnarray}.

\vspace{.12in} \noindent \textbf{4.5 Classical Hamiltonian mechanics
and traditional Hilbert space quantum mechanics as subdisciplines of
supmech}

\vspace{.1in} A classical Hamiltonian system $(P, \omega_{cl},
H_{cl})$ [where P is the phase space which is a symplectic manifold
with the classical symplectic form $\omega_{cl} \equiv \sum
dp_{\alpha} \wedge dq^{\alpha}$ (in canonical coordinates) and
$H_{cl}$ is the classical Hamiltonian, a smooth real-valued function
on P] is a special case of a supmech Hamiltonian system
$(\sca, \sone(\sca), \omega, H)$ with $\sca = \sca_{cl}
\equiv C^{\infty}(P; \mathbb{C}), \ \sone(\sca) = P$ (Dirac measures on the
phase space P identified with points of P), \ $ \omega = \omega_{cl}$
and H = $H_{cl}$; the supmech PBs are now the traditional classical
PBs. The supmech Hamilton's equation (10) is now the classical
Hamilton's equation. Representing states by probability densities in
phase space, Eq.(12) goes over, in appropriate cases
(for $ P = \mathbb{R}^{2n}$, for example, after the obvious partial
integrations),  to the classical Liouville equation for the density
function. The CC condition can be easily verified in this case
(II, section 2.2). The supmech events are now the
characteristic/indicator functions corresponding to the
Borel subsets of P (which correspond to events in classical
probability theory) (II, section 2.1).

To see the traditional Hilbert space QM as a subdiscipline of
supmech, it is useful to introduce the concept of a \emph{quantum
triple} $(\sch, \mathcal{D}, \sca)$ where \sch \ is a complex
separable Hilbert space, $\mathcal{D}$ a dense linear subset of \sch
\ and \sca \ an Op$^*$-algebra of operators based on
($\sch,\mathcal{D}$). [Such an algebra is a family  of operators
which, along with their adjoints, map \scd \ into itself. The
*-operation on the algebra is defined as the restriction of the
Hilbert space adjoint on $\mathcal{D}$. These are the algebras of
operators (not necessarily bounded) appearing in the traditional
Hilbert space QM; for example, the operator algebra  generated by
the position and momentum operators in the Schrodinger
representation for a nonrelativistic spinless particle  (the
Heisenberg algebra) belongs to this class, with $\sch =
L^2(\mathbb{R}^3)$ and $\mathcal{D} = \mathcal{S}(\mathbb{R}^3)$.]

Here we shall consider only the \emph{standard quantum triples} by
which we mean those in which (i) the algebra \sca \ is special in
the sense described above, and (ii) \sca \ acts irreducibly on
($\sch,\mathcal{D}$) [i.e. there does not exist a smaller quantum
triple $(\sch^{\prime}, \mathcal{D}^{\prime}, \sca)$ with
$\mathcal{D}^{\prime} \subset \mathcal{D}, \ \sca
\mathcal{D}^{\prime} \subset \mathcal{D}^{\prime}$ and
$\sch^{\prime}$ is a proper subspace of \sch]. The quantum triple
associated with the Schr$\ddot{o}$dinger representation for a
non-relativistic spinless particle mentioned above satisfies these
conditions.

With \sca \ special, one can define the quantum symplectic form
$\omega_Q = -i \hbar \omega_c$ which gives the Poisson brackets of
Eq.(9). With the \sca-action irreducible, the space $\sone(\sca)$ of
pure states of \sca \ consists of vector states corresponding to
normalized vectors in $\mathcal{D}$. Choosing an appropriate self
adjoint element H of \sca \ as the Hamiltonian operator, we have a
quantum Hamiltonian system $(\sca, \sone(\sca), \omega_Q, H)$  as a
special case of a supmech Hamiltonian system. The PObVMs are now the
traditional POVMs (positive operator-valued measures). It was shown
in II that the Born probabilities in traditional quantum mechanics
can always be expressed in the form (5).

With the quantum PBs of Eq.(9), the supmech Hamilton's equation
(10) goes over to the traditional Heisenberg equation of motion.
General states are represented by density operators $\rho$
satisfying the condition $|Tr(\overline{\rho A})| < \infty$ for
all observables A in \sca \ (where the overbar indicates closure
of the operator). The CC condition holds in this case as well
(II, section 2). Noting that $Tr(\overline{\rho_1 A}) =
Tr ({\overline{\rho_2 A}})$ for all $A \in \sca$ implies
$\rho_1 = \rho_2$, Eq.(12) goes over to the von Neumann equation
\begin{eqnarray} \frac{d\rho(t)}{dt} = (-i\hbar)^{-1}[\rho(t), H].
\end{eqnarray}

\vspace{.12in} \noindent \textbf{4.6 Quantum-classical
correspondence}

\vspace{.1in} This feature of supmech (of accommodating both
classical and quantum mechanics) facilitates a transparent treatment
of quantum-classical correspondence. The strategy adopted in II was
to start with a quantum Hamiltonian system, transform it to an
isomorphic supmech Hamiltonian system involving phase space
functions and $ \star $-products (Weyl-Wigner-Moyal formalism) and
show that, in this latter Hamiltonian system, the subclass of phase
space functions in the system algebra which go over to smooth
functions in the $\hbar \rightarrow 0$ limit yield the corresponding
classical Hamiltonian system. The working of this strategy was
demonstrated for the case of a spinless nonrelativistic particle. It
was, however, clear that the treatment permitted  trivial
generalization to systems with phase space $\mathbb{R}^{2n}$. We
collect below the $\mathbb{R}^{2n}$-analogues of some equations from
section 4 of II. [The integrals in equations (20-26) below are over
$\mathbb{R}^{n}$.]

Given a quantum triple $(\sch, \mathcal{D}, \sca)$ where $\sch = L^2
(\mathbb{R}^{n}), \mathcal{D} = \mathcal{S}(\mathbb{R}^{n})$ and
\sca \ an $Op^*$-algebra based on $(\sch, \mathcal{D})$, we have,
for any $A \in \sca$ and $ \phi,\psi $ normalized elements in
$\mathcal{D}$,
\begin{eqnarray}
(\phi,A\psi) = \int \int \phi^*(y) K_{A}(y,y^\prime) \psi(y^\prime)
dydy^\prime
\end{eqnarray}
where the kernel $ K_A$ is a (tempered) distribution. The Wigner
function $A_W$ corresponding to A is defined as the function on
${R}^{2n}$ given by \begin{eqnarray} A_W(x,p) = \int
exp[-ip.y/\hbar]K_A(x + \frac{y}{2}, x - \frac{y}{2})dy.
\end{eqnarray}
Given a density operator $\rho$ on \sch \ such that
$|Tr (\overline{A \rho})| < \infty$ for all $A \in \sca$ and
defining $\rho_W$ as above, we have
\begin{eqnarray} Tr(\overline{A \rho})
= \int \int A_W(x,p) \rho_W(x,p)dxdp. \end{eqnarray}
The Wigner function $\rho_W$ is real but generally
not non-negative.

 Introducing, in $ \mathbb{R}^{2n},$ the notations
 $ \xi $ = (x,p), $ d\xi = dxdp $ and $ \sigma ( \xi, \xi^{'} )
 = p.x^{'} - x.p^{'} $ (the symplectic form in
 $ \mathbb{R}^{2n} $), we have, for  A,B $ \in \sca $
\begin{eqnarray}
(AB)_{W}(\xi) & =  &     (2\pi)^{-6} \int \int  exp [ -i \sigma
                               (\xi - \eta, \tau)] A_{W} ( \eta +
                               \frac{\hbar \tau}{4} ). \nonumber \\
& \ & . B_{W} (\eta - \frac{\hbar \tau }{4})d\eta d\tau \nonumber \\
                        &  \equiv &  (A_{W} \star B_{W}) ( \xi).
\end{eqnarray}
The associativity condition $ A(BC) = (AB)C $ implies the
corresponding condition $ A_W \star (B_W \star C_W) = (A_W \star
B_W) \star C_W $ in the space $ \sca_W$ of the Wigner functions
corresponding to the elements of \sca \  which is a complex
associative non-commutative, unital
*-algebra (with the \emph{star-product} of Eq.(23) as product and
complex conjugation as involution) isomorphic (as a star-algebra) to
\sca. Under this isomorphism, the quantum symplectic form $\omega_Q
= -i \hbar \omega_c$ on \sca \ goes over to the 2-form $\omega_W =
-i \hbar \omega_c^W$ where $\omega_c^W$ is the canonical form on
$\sca_W$; this makes the pair $(\sca_W, \omega_W )$ a symplectic
algebra isomorphic to $(\sca, \omega_Q)$. The corresponding PB on
$\sca_W$ is given by the \emph{Moyal bracket}
\begin{eqnarray}
\{ A_{W}, B_{W} \}_{M} \equiv (-i\hbar)^{-1} ( A_{W} \star B_{W} -
B_{W} \star A_{W} ).
\end{eqnarray}

For  functions f, g in $\mathcal{A}_{W} $ which are smooth and such
that   $ f(\xi)$ and $ g(\xi)$ have no $ \hbar-$dependence, we have,
from Eq.(23),
\begin{eqnarray}
f \star g = fg - (i\hbar/2) \{ f, g \}_{cl} + O ( \hbar^{2} ).
\end{eqnarray}
The functions $ A_{W} (\xi) $ will have, in general, some $ \hbar $
dependence and the $ \hbar \rightarrow 0 $ limit may be singular for
some of them. We denote by $(\mathcal{A}_{W})_{reg}$ the subclass of
functions in $ \mathcal{A}_{W} $ whose $ \hbar \rightarrow 0 $
limits exist and are smooth (i.e. $ C^{\infty} $ ) functions;
moreover, we demand that the Moyal bracket of every pair of
functions in this subclass also have smooth limits. This class is
easily seen to be a subalgebra of $ \mathcal{A}_{W}$ closed under
Moyal brackets.  Now, given two functions $A_W$ and $B_W$ in this
class, if $ A_{W} \rightarrow A_{cl} $  and $ B_{W} \rightarrow
B_{cl} $  as $ \hbar \rightarrow 0,$ then  $ A_{W}  \star B_{W}
\rightarrow A_{cl} B_{cl} $; the subalgebra
$(\mathcal{A}_{W})_{reg}$, therefore, goes over, in the $ \hbar
\rightarrow 0 $ limit , to a subalgebra $\mathcal{A}_{cl}$ of    the
commutative algebra $ C^{\infty}(\mathbb{R}^{2n}, \mathbb{C}) $ (with
pointwise product as multiplication). The Moyal bracket of Eq.(24)
goes over to the classical PB $\{ A_{cl}, B_{cl} \}_{cl}$; the subalgebra
$\mathcal{A}_{cl}$, therefore, is closed under the classical Poisson
brackets. The classical PB $\{, \}_{cl}$ determines the classical
symplectic form $\omega_{cl}$. In the $\hbar \rightarrow 0$ limit,
therefore, we have the classical symplectic algebra $(\sca_{cl},
\omega_{cl})$. In situations where $H_W \in (\sca_W)_{reg}$ admitting the
$\hbar \rightarrow 0$ limit $H_{cl}$, we have, in this limit, the
classical Hamiltonian system $(\sca_{cl}, \omega_{cl}, H_{cl}).$ [This
is the case, for example, for the Hamiltonian operator $ H =
(2m)^{-1}\mathbf{P}^2 + V(\mathbf{X})$ with V a smooth function.]

When the $\hbar \rightarrow 0$ limits of $A_W$ and $\rho_W$ on the
right hand side of Eq.(22) exist (call them $A_{cl}$ and
$\rho_{cl}$), we have, in this limit,
\begin{eqnarray} Tr(\overline{A \rho}) \rightarrow \int \int A_{cl}(x,p)
\rho_{cl}(x,p) dxdp. \end{eqnarray} The quantity $\rho_{cl}$ can  be
shown to be non-negative (and, therefore, a genuine density function
on the phase space $\mathbb{R}^{2n}$).

We shall make, in our treatment of measurements below, the fairly
safe assumption that this strategy works for the apparatus treated
as a quantum system. [See the axiom A8(b,c) in section 8.] This will
enable us to exploit the fact that the apparatus admits a classical
description to a very good approximation.

\vspace{.15in} \noindent \textbf{\large 5. Treatment of a Quantum Measurement
in Supmech}

\vspace{.1in} We shall now treat the (S +A) system in the framework
of section IV D above treating both, the system S and the apparatus
A, as quantum Hamiltonian systems. Given the two quantum triples
$(\sch_S, \mathcal{D}_S, \sca_S)$ and $(\sch_A, \mathcal{D}_A,
\sca_A)$ corresponding to S and A, the quantum triple corresponding
to (S+A) is $(\sch_S \otimes \sch_A, \mathcal{D}_S \otimes
\mathcal{D}_A, \sca_S \otimes \sca_A )$.

 A general pointer observable for A is of the form
\begin{eqnarray}
J = \sum_j b_j P_j
\end{eqnarray}
where $P_j$ is the projection operator onto the space of states in
$\sch_A$ corresponding to the pointer position $M_j$ [considered as
an apparatus property; for a detailed treatment of the relationship
between classical properties and quantum mechanical projectors, see
(Omnes [6,8]) and references therein] and $b_j$s are real
numbers such that $b_j \neq b_k$ for $j \neq k$. In purely quantum
mechanical terms, the projector $P_j$ represents the question (von
Neumann [3]; Jauch [5]) : `Is the pointer at position $M_j$?' The
observable J has different `values' at different pointer positions.
Since one needs only to distinguish between different pointer
positions, any observable J of the above mentioned specifications
can serve as a pointer observable.

The phase space function $P^W_j$ corresponding to the projector
$P_j$ is  supposedly approximated well by a function $P_j^{cl}$ on
the phase space $\Gamma$ of the apparatus A (the $\hbar \rightarrow
0$ limit of $P^W_j$). Now, in $\Gamma$, there must be
non-overlapping domains $D_j$ corresponding to the pointer positions
$M_j$. In view of the point (iv) in section III, different points in a
single domain $D_j$ are not distinguished by the experiment. We can,
therefore, take $P_j^{cl}$ to be proportional to the
characteristic/indicator function $\chi_{D_j}$ of the domain $D_j$;
it follows that the phase space function $J^W$ corresponding to the
operator J above is approximated well by the classical pointer
observable
\begin{eqnarray} J^{cl} = \sum_j b_j^{\prime} \chi_{D_j}
\end{eqnarray} where $b_j^{\prime}$s have properties similar to the
$b_j$s above.

The pointer states $\phi_j^{(A)}$ corresponding to the  pointer
positions $M_j$ are represented by density operators $\rho^{(A)}_j$
supposedly such that

\noindent (i) $Tr (\overline{\phi_j^{(A)} P_k}) = \delta_{jk}$;

\noindent (ii) the phase space functions $\rho^{(A)W}_j$
corresponding to them are approximated well by the classical phase
space density functions $\rho_j^{(A)cl} $ which vanish outside the
domain $D_j$.

We shall take $ H_{int} = F \otimes K$ (absorbing the coupling
constant in K) where F is the measured quantum observable and K is a
suitably chosen apparatus observable . We shall make the usual
assumption that, during the measurement interaction, $H_{int}$ is
the dominant part of the total Hamiltonian $(H \simeq H_{int})$. The
unitary operator U of section II describing the measurement
interaction in the von Neumann scheme is now proposed to be replaced
by the measurement operator M in supmech which implements the
appropriate canonical transformation on the states of the (S +A)
system. It is  given by $ M \equiv exp[\tau \tilde{\partial}_H]$
where $\tau = t_f-t_i$ is the time interval of measurement
interaction and $\tilde{\partial}_H$ is the evolution generator in
the supmech Liouville equation [see Eq.(13)].

Assuming, again, that the measurement is ideal and denoting the
`ready state' of the apparatus by $\phi_0^{(A)}$, we have the
following  analogue of Eq.(1):
\begin{eqnarray}
M (|\psi_j><\psi_j| \otimes \phi_0^{(A)}) = |\psi_j><\psi_j| \otimes
\phi_j^{(A)}.
\end{eqnarray}
Here and in the following developments, we have identified the
quantum states of the system S with the corresponding density
operators. When the system is initially in the superposition state
$|\psi>$ as in section II, the initial and final (S+A)- states are
\begin{eqnarray}
\Phi_{in} = |\psi><\psi| \otimes \phi_0^{(A)}; \ \ \Phi_f =
M(\Phi_{in}).
\end{eqnarray}

Note that the `ready' state may or may not correspond to one of the
pointer readings. (In a voltage type measurement, it does; in the
Stern-Gerlach experiment with spin half particles, it does not.) For
the assignment of the $\Gamma$-domain to the `ready' state, the
proper interpretation (which covers both the situations above) of
the ready state is `not being in any of the (other) pointer states'.
Accordingly, we assign, to this state, the domain
\begin{eqnarray}
\tilde{D}_0 \equiv \Gamma - \cup_{j \neq 0}D_j
\end{eqnarray}
where the condition $j \neq 0$ on the right is to be ignored when
the `ready' state is not a pointer state.

We must now take care of the point (iii) of section 3. When the
measurement interaction is over, the apparatus, left to itself, will
quickly occupy, in any single experiment, a pointer position $M_j$
(depending on the region of the phase space $\Gamma$  it happens to
be in after the measurement interaction). For the ensemble of (S +A)
systems described by the initial state $\Phi_{in}$, the final state
(after `settling down') must be of the form
\begin{eqnarray} \hat{\Phi}_f = \sum_j p_j \rho^{(S)}_j \otimes
\phi_j^{(A)} \end{eqnarray} where $\rho_j^{(S)}$ are some states of
S. Eq.(32) incorporates the net effect of the processes involved in
the `settling down' of the apparatus. The unknowns $p_j$ and
$\rho^{(S)}_j$ must be determined by identifying the conditions that
must be satisfied by the processes involved in the above mentioned
`settling down'.

During the transition from the state $\Phi_f$ to $\hat{\Phi}_f$, the
change taking place in the system (S+A) is predominantly `settling
down' of the apparatus which, in view of the stability property
(iii) above, is not expected to change the expectation value of a
pointer observable J. We must have, therefore,
\begin{eqnarray} \hat{\Phi}_f (A \otimes J) = \Phi_f( A \otimes J)
\end{eqnarray}
for all system observables A and all pointer observables J of the
form (27). It is this condition, based on physical reasoning,
which replaces von Neumann's projection postulate in our treatment.

Now, $\Phi_f = \Phi_f^{\prime} + \Phi_f^{\prime \prime}$ where
\begin{eqnarray}
\Phi_f^{\prime} & = & M \left( \sum_j |c_j|^2 [|\psi_j><\psi_j|
\otimes
              \phi_0^{(A)}] \right) \nonumber \\
& = & \sum_j |c_j|^2 [|\psi_j><\psi_j| \otimes
                       \phi_j^{(A)}] \end{eqnarray}
(where we have used the fact that the  mapping
$M \equiv exp[\tau \tilde{\partial}_H]$ on states preserves convex
combinations) and
\begin{eqnarray} \Phi_f^{\prime \prime} = M \left( [\sum_{j \neq k}
c_k^* c_j |\psi_j><\psi_k|] \otimes  \phi_0^{(A)} \right) \equiv M
(R).
\end{eqnarray}
[Note that R, the operand of M, is not an (S +A)-state; here M has
been implicitly extended by linearity to the dual space of the
algebra $\sca_S \otimes \sca_A$.]

We shall now prove that \begin{eqnarray} W \equiv \Phi_f^{\prime
\prime} (A \otimes J ) \simeq 0. \end{eqnarray}

\vspace{.1in} \noindent \emph{Proof.} Transposing the M operation to
the observables and adopting the phase space description of the
apparatus, we have
\begin{eqnarray}
W & = & <exp(\tau \tilde{\partial}_H )(R), A \otimes J > \   =
\ <R, [exp(\tau \partial_H) (A \otimes J)] \nonumber \\
& = & < (\sum_{j \neq k} c_k^* c_j |\psi_j><\psi_k|) \otimes
\phi_0^{(A)}, [exp(\tau \partial_H)] (A \otimes J)> \nonumber \\
  & = & \int_{\Gamma} d \Gamma \rho^{(A)W}_0 \sum_{j \neq k}
        c_k^* c_j < |\psi_j><\psi_k|,
exp(\tau \partial_{H^{\prime}})
        (A \otimes J^W)>
\end{eqnarray}
where  $d \Gamma$ is the phase space volume element, $\rho^{(A)W}_0$
is the Wigner function corresponding to the state $\phi^{(A)}_0$ and
$H^{\prime} = F\otimes K^W$ [see Eq.(22)]. Using equations (9), (17)
and (24) above, we have
\begin{eqnarray}
\partial_{H^{\prime}}(A \otimes J^W)
& = & \{ F \otimes K^W, A \otimes J^W \} \nonumber \\
& = & (-i\hbar)^{-1} \left( [F,A] \otimes \frac{K^W *
                      J^W + J^W * K^W}{2} \right.\nonumber \\
& \ & + \left. \frac{FA + A F}{2}
                      \otimes (K^W * J^W - J^W * K^W) \right).
\end{eqnarray}
Given the fact that the apparatus is well described classically, we
have $K^W \simeq K^{cl}$ and $J^W \simeq J^{cl}$ to a very good
approximation. This gives
\[ \partial_{H^{\prime}} (A \otimes J^W) \simeq
(-i\hbar)^{-1} K^{cl} J^{cl} [F,A] \] which, in turn, implies
(recalling the notation $ D_F (A) = [F,A]$)
\begin{eqnarray*} < |\psi_j><\psi_k|&,& exp (\tau
\partial_{H^{\prime}}) (A \otimes J^W)> \\ & = & < |\psi_j><\psi_k|,
\ exp (\frac{i \tau}{\hbar} K^{cl} D_F) (A)> J^{cl} \\
& = & <\psi_k |exp (\frac{i \tau}{\hbar} K^{cl} D_F) (A) |\psi_j>
J^{cl} \\ & = & exp[\frac{i \tau}{\hbar} K^{cl}(\lambda_k -
\lambda_j)] <\psi_k| A |\psi_j> J^{cl}. \end{eqnarray*}
 We now have, replacing, in Eq.(37), $\rho^{(A)W}_0$ by its classical
approximation $\rho_0^{(A)cl}$,
\begin{eqnarray}
W \simeq  \int_{\tilde{D}_0} d \Gamma \rho^{(A)cl}_0
    \sum_{j \neq k} c_k^* c_j \ exp[\frac{i}{\hbar}
    (\lambda_k - \lambda_j)K^{cl}\tau] J^{cl} <\psi_k|A|\psi_j>.
\end{eqnarray}
Let
\begin{eqnarray} <K^{cl}>_0 \  \equiv  \int_{\tilde{D}_0} K^{cl}
\rho_0^{(A)cl} d \Gamma \end{eqnarray} (the mean value of $K^{cl}$
in the domain $\tilde{D}_0$; we shall give an argument below
showing that it is nonzero). Putting $K^{cl} = <K^{cl}>_0 s$,
taking s to be one of the integration variables and writing $d
\Gamma = ds d\Gamma^{\prime}$, we have
\begin{eqnarray} W \simeq \int_{\tilde{D}_0} ds d\Gamma^{\prime}
\rho_0^{(A)cl} \sum _{j \neq k} c_k^*c_j \
exp[\frac{i}{\hbar}\eta_{jk}s]J^{cl} <\psi_k|A|\psi_j>
\end{eqnarray} where
\begin{eqnarray} \eta_{jk} = (\lambda_k - \lambda_j)<K^{cl}>_0 \tau.
\end{eqnarray}
Note that s is a real dimensionless variable with a bounded domain
of integration [see remark (iii) below].

We shall now argue that, for $j \neq k,$
\begin{eqnarray}  | \eta_{jk}| > > \hbar. \end{eqnarray}
[This is not obvious; when F is a component of spin, for example,
the quantity $(\lambda_k - \lambda_j)$ is a scalar multiple of
$\hbar$.] To this end, we invoke the apparatus feature (v) of
section III. A reasonable procedure for  formulating a criterion for
macroscopic distinguishability of different pointer positions would
be to identify a quantity of the dimension of action which could be
taken as characterizing the physical separation between two
different pointer positions and show that its magnitude is much
larger than $\hbar$. The objects $\eta_{jk}$ (for $j \neq k$) are
quantities of this type. A simple way of seeing this is to treat
Eq.(43) as the time-energy uncertainty inequality $ |\Delta E \Delta
t| > > \hbar $ where $ \Delta t = \tau$ and $\Delta E$ is the
difference between the energy values corresponding to the apparatus
locations in two different domains $D_j$ and $D_k$ in $\Gamma$.
Recalling that $H \simeq H_{int}$ during the relevant time interval,
we have
\begin{equation}
\Delta E \simeq (\lambda_k - \lambda_j)<K^{cl}>_0.
\end{equation}
[See the remark (ii) below.] The inequality (43) then follows from
the assumed macroscopic distinguishability of different pointer
positions. This assumption along with the argument above also
implies $<K^{cl}>_0 \neq 0$ as promised above.

The large fluctuations implied by Eq.(43) wipe out the integral in
Eq.(41) giving $ W \simeq 0 $ as desired. $ \Box$

\vspace{.1in} \noindent \emph{Remarks}. (i) For an argument,
starting from the condition of macroscopic distinguishability of
pointer positions and arriving at the time-energy uncertainty
inequality in the context of the Stern-Gerlach experiment, see
(Gottfried [15]).

\vspace{.1in} \noindent (ii) How does one justify the appearance of
the mean value of $K^{cl}$ in the `ready' state in the expression
for $\Delta E$ in the energy time inequality above ? A plausible
answer is this : Since the apparatus is initially in the `ready'
state and since K appears in $H_{int}$,, it is the quantity
$<K^{cl}>_0$ which will, at the classical level,  be effective in
determining the  probabilities of transitions to the various domains
$D_j$.

\noindent [A more refined argument : Suppose, at time $t= t_i$, the
system point of the apparatus A, considered as a classical system,
in the phase space $\Gamma$ is $\xi_0 \in \tilde{D}_0$. With the
system in the initial state $|\psi_k>$, the effective classical
Hamiltonian is $H^{cl(k)}= \lambda_k K^{cl}$. After the measurement
interaction, at $t= t_f$,  we have the system point of A at
$\xi(t_f) = \xi_{0k} \in D_k$. For the quantity $\Delta E$
considered above, a good estimate is \begin{equation} \Delta E
\simeq \int d \Gamma(\xi_0) \rho^{(A)cl}_0(\xi_0) [\lambda_k
K^{cl}(\xi_{0k}) - \lambda_j K^{cl}(\xi_{0j})]. \end{equation} But
the Hamiltonians $H^{cl(k)}$ and $H^{cl(j)}$ conserve the quantity
$K^{cl}$. This gives $K^{cl}(\xi_{0k}) = K^{cl}(\xi_{0})
 = K^{cl}(\xi_{0j})$, hence Eq.(44).]

\vspace{.1in} \noindent (iii) Physical quantities related to the
apparatus must, in their classical description, be bounded functions
on $\Gamma$. (Even observables like the Cartesian components of
position or momentum of macroscopic parts/components of the
apparatus must vary in finite intervals.) Boundedness of the domain
of integration of the variable s now follows from the relation
$|s|_{max} = |(<K^{cl}>_0)^{-1} K^{cl}|_{max}$. [\emph{Note.} In the
example in the next section, the dimensionless variable u, playing
the same role as s here, has domain of variation of length of order
1. In the general case, let $ s_1 \leq s \leq s_2$. If $ s_2 - s_1
\leq  2 \pi$, no further argument is necessary. If $ s_2 - s_1
> 2 \pi$, put $ s = (s_2 - s_1) w$; now the integration variable w
is similar to u and the additional factor $(s_2 - s_1)$ in the
exponent is welcome.]

\vspace{.1in} Equations (33), (32) and (36) now give
\begin{eqnarray*} 0 & \simeq &
(\hat{\Phi}_f - \Phi_f^{\prime}) ( A \otimes J ) \\
& = & \sum_j \phi_j^{(A)} (J) Tr (\overline{[p_j \rho^{(S)}_j - |c_j|^2
|\psi_j><\psi_j|]A}) \\
& = & \sum_j b_j Tr (\overline{[p_j \rho^{(S)}_j - |c_j|^2
|\psi_j><\psi_j|]A})
\end{eqnarray*} which must be true for  arbitrary $b_j$ in Eq.(27)
satisfying the stated condition. This gives
\[ Tr(\overline{[ p_j \rho^{(S)}_j - |c_j|^2 |\psi_j><\psi_j|]A}) =
0 \] for all j and all system observables A and, therefore, for all j,
\[ p_j \rho^{(S)}_j = |c_j|^2 |\psi_j><\psi_j|. \]
Finally, therefore, we have $\hat{\Phi}_f = \Phi_f^{\prime} $ which
is precisely the state obtained from $\Phi_f$ by applying the von
Neumann projection.

This completes the derivation of the von Neumann projection rule.
This has been obtained through straightforward physics; there is no
need to give any ad hoc prescriptions. The derivation makes it clear
as to the sense in which this reduction rule should be understood :
it is a prescription to correctly take into consideration the effect
of the `settling down' of the apparatus after the measurement
interaction for obtaining the final state of the system (S + A)
observationally constrained as  in items (iv) and (v) of section 3.

Eq.(41), followed by the reasoning above, represents,  in a
\emph{live} form, the operation of environment-induced decoherence.
To see this, note that, the domain $\tilde{D}_0$  may be taken to
represent the internal environment of the apparatus. With this
understanding, the mechanism wiping out the unwanted quantum
interference terms is, indeed, the environment-induced decoherence.
In the treatment presented here this mechanism becomes automatically
operative. (Even the external environment can be trivially included
by merely saying that the system A above represents `the apparatus
and the external environment'.)

\vspace{.15in} \noindent \textbf{\large 6. Example : The Stern-Gerlach
Experiment}

\vspace{.1in} As an illustration of the automatic appearance of the
decoherence mechanism in the supmech based treatment of quantum
measurements presented in the previous section, we consider the
Stern-Gerlach experiment (Busch, Grabowski and Lahti [16]; Omnes [6];
Gottfried [15]; Cohen- Tannoudji, Diu  and Lalo$\ddot{e}$
[17]) with, say, silver atoms (which means spin s = $\frac{1}{2}$).
A collimated beam of (unpolarized) silver atoms is made to pass
through inhomogeneous magnetic field after which the beam splits
into two beams corresponding to atoms with $S_z = \pm
\frac{\hbar}{2}$. The spin and magnetic moment operators of an atom
are $\mathbf{S} = \frac{\hbar}{2} \mathbf{\sigma}$ and $\mathbf{\mu}
= g \mathbf{S}$ (where g is the magnetogyric ratio). Let the
magnetic field be $\mathbf{B}(\mathbf{r}) = B(z) \mathbf{e_3}$ (in
obvious notation). [Refinements (Potel et al [18]) introduced to
ensure the condition $ \mathbf{\bigtriangledown.B} =0$ do not affect
the essential results obtained below.] We have
\begin{eqnarray}
H_{int} = -\mathbf{\mu . B} = -g B(z) S_3.
\end{eqnarray}
The force on an atom, according to Ehrenfest's theorem, is
\begin{eqnarray}
\mathbf{F} = - \mathbf{\bigtriangledown} < - \mathbf{\mu . B}> = g
\frac{d B(z)}{dz}<S_3> \mathbf{e_3}
\end{eqnarray}
where the average is taken in the quantum state of the atom. During
the experiment, the internal state of the atom remains unchanged (to
a very good approximation); only its center of mass \textbf{r} and
spin \textbf{S} have significant dynamics. In this experiment, $S_3$
is the measured quantum observable  and \textbf{r} acts as the
operative apparatus variable.

\begin{sloppypar}Let us assume that the beam initially moves in the
positive x-direction, the pole pieces are located in the region $x_1
\leq x \leq x_2$ and the detectors located in the plane $ x = x_3 >
x_2$ (one each in the regions $z >0$ and $z < 0;$ these regions
contain the emergent beams of silver atoms corresponding,
respectively, to $S_3 = +\frac{\hbar}{2}$ and $S_3 = -
\frac{\hbar}{2}$). We have, in the notation used above, $F = S_3$
and K = - g B(z). Assuming the experiment to start when the beam
reaches at $ x = x_1$, the phase space of the apparatus is
\begin{eqnarray}
\Gamma = \{ (x, y, z, p_x, p_y, p_z) \in \mathbb{R}^{6}; x \geq x_1,
* \}
\end{eqnarray}
where * indicates the restriction that, for $x_1 \leq x \leq x_2, $
the space available for the movement of atoms is the one between the
two pole pieces. For the order of magnitude calculation below, we
shall ignore the shape of the pole pieces and take * to imply $z_1
\leq z \leq z_2$. \end{sloppypar}

The domains $D_1$ and $D_2$ corresponding to the two pointer
positions are
\begin{eqnarray*}
D_1 & = & \{ (x, y, z, p_x, p_y, p_z) \in \Gamma; x > x_2, p_z >0 \} \\
D_2 & = & \{ (x, y, z, p_x, p_y, p_z) \in \Gamma; x > x_2, p_z < 0
\};
\end{eqnarray*}
the domain $\tilde{D}_0 = \Gamma -(D_1 \cup D_2)$. For simplicity,
let us take $ B(z) = b_0 + b_1 z$ where $b_0$ and $b_1$ are
constants. For $j \neq k$, we have $\lambda_j - \lambda_k = \pm
\hbar$. The relevant integral is [see Eq.(41) above]
\begin{eqnarray}
I = \int_{z_1}^{z_2} dz (...)exp[\pm \frac{i}{\hbar}\mu b_1 z \tau]
\end{eqnarray}
where $\mu = g \hbar$. Putting $z = (z_2 - z_1) u$, the new
integration variable u is a dimensionless variable taking values in
a domain of length of order one. The quantity of interest is
\begin{eqnarray}
|\eta| = \mu |b_1| (z_2 - z_1) \tau.
\end{eqnarray}
According to the data in (Cohen-Tannoudji, Diu  and Lalo$\ddot{e}$
[17]) and (Goswami [19]; problem 4.6), we have ($v_x$ is the x-
component of velocity of  the silver atom)
\begin{eqnarray*}
|b_1| \sim |\frac{dB}{dz}| \sim 10^5 \ gauss/cm \\
z_2 - z_1 \simeq 1 \ mm, \ \ v_x \sim 500 \ m/sec \\
x_2- x_1 = 3 \ cm, \ \ x_3 - x_2 = 20 \ cm
\end{eqnarray*}
This gives
\begin{eqnarray*}
\tau \sim \frac{x_3 - x_1}{v_x} \sim 5 \times 10^{-4} sec.
\end{eqnarray*}
Denoting the Bohr magneton by $\mu_b$ and putting $ \mu \sim \mu_b
\simeq 0.9 \times 10^{-20}$ erg/gauss, we have $|\eta| \sim
10^{-19}$erg-sec. With $ \hbar \simeq 1.1 \times 10^{-27}$ erg-sec,
we have, finally, $(|\eta|/\hbar) \sim 10^8$, confirming the strong
suppression of the undesirable quantum interferences.

\vspace{.15in} \noindent \textbf{\large 7. Comparison with the Traditional
Decoherence Program}

\vspace{.12in} In the traditional  decoherence program,
one invokes the interaction of the system (S + A) with the environment
(to be denoted as $\mathcal{E}$), treat the combined system
(S + A + $\mathcal{E}$) as in section 2 up to the pre-measurement
level of Eq.(2) and then take the partial trace of the density operator
of (S + A + $\mathcal{E}$) over the environment $\mathcal{E}$ to obtain
the reduced density operator for the system (S + A). Making some
plausible assumptions about the states of the environment (which can
be verified in concrete model calculations), one can show that the
reduced density operator for (S+A) has the desired form (4) incorporating the
von Neumann projection. Taking trace over $\mathcal{E}$ is interpreted
as ignoring uncontrolled and unmeasured degrees of freedom; this is,
supposedly, similar to the procedure of deriving the probability
$\frac{1}{2}$ for `heads' as well for `tails' in the experiment of
tossing a fair coin by averaging over the uncontrolled and unmeasured
degrees of freedom of the environment of the coin.

A critical look at these developments (see Bub [10], Adler [11]), however,
shows that there are loopholes. As argued by Bub, the two averaging procedures
are not on the same footing. In the coin toss experiment, when, ignoring the
environment, we claim that, the probability of getting `heads' in a particular
toss of the coin is $\frac{1}{2}$, we can also claim that we do, in fact, get
\emph{either} `heads' or `tails' on each particular toss. A definite outcome
can be predicted if we take into consideration the environmental effects and
details of initial conditions of the toss. In the treatment of a quantum
measurement in the traditional decoherence program as outlined above, however,
we cannot claim that, taking the environment into consideration, a definite
outcome of the experiment will be predicted. In fact, taking the environment
will give us back a troublesome equation of the form (2) [with A replaced by
(A + $\mathcal{E}$)] which is obtined in a von Neumann type treatment of the
system (S + A + $\mathcal{E}$).

The problem really lies with some unsatisfactory features and inadequacies
in the treatment of the apparatus in von Neumann's treatment of measurements.
(See section 3.) In the present work, the physics of the apparatus has been
taken seriously and the points mentioned in section 3 have been properly
incorporated in the supmech based treatment of measurements in section 5. The
decoherence effects arise because  observations on the apparatus are
restricted to macroscopically distinguishable pointer readings which
inevitably leads to averaging over the inoperative degrees of freedom
(internal environment) [see Eq.(41).] Here is an `averaging over uncontrolled
and unmeasured degrees of freedom' which is arising in a consistent scheme of
mechanics; there are obviously no problems of consistency or logical
coherence here.

Another important positive feature of the present treatment is that the
paradoxical aspect of the `Heisenberg cut' or Bohr's `quantum-classical divide'
has been gotten rid of in the most natural way  --- by treating the apparatus
carefully as a quantum system approximated well by a classical one.

\vspace{.15in} \noindent \textbf{\large 8. The Eighth Axiom of the
Supmech Program}

\vspace{.1in} A provisional set of seven axioms, underlying the
plan to do `all physics' in the framework of a noncommutative
symplectic geometry based universal mechanics adopted in this series
of papers (\emph{the supmech program}), was presented in section 5
of II. The use of the word `provisional' reflects the expectation
that a stage may come when, after achieving some successes, a more
compact set of axioms (which may themselves be `provisional' at a
higher level) is found more suitable. Till such a stage is reached,
the list of provisional axioms is expected to increase with
additions in coverage of the program. The new assumptions made in
the present work are being listed below as the eighth provisional
axiom.

\vspace{.12in} \noindent \textbf{A8.} \emph{Measurements}. In a
measurement involving a `measured system' S and apparatus A, the
following items hold :

\vspace{.1in} \noindent (a) Both S and A are standard quantum
systems (as defined in section 3.2 of II; it means that that the
system algebra is a noncommutative $\star$-algebra generated by
a finite number of fundamental observables and the unit element).

\vspace{.1in} \noindent (b) The supmech Hamiltonian system
$(\sca^{(A)}, \sone^{(A)}, \omega^{(A)}, H^{(A)})$ corresponding to
the apparatus admits an equivalent (in the sense of section IV C)
phase space realization (in the Weyl-Wigner-Moyal scheme)
$(\sca_W^{(A)}, \mathcal{S}_{1W}^{(A)}, \omega_W^{(A)}, H_W^{(A)})$.

\vspace{.1in} \noindent (c) Elements of $\sca_W^{(A)}$ and
$\mathcal{S}^{(A)}_{1W}$ appearing in the description of dynamics of
the coupled system (S+A) admit $\hbar \rightarrow 0$ limits and are
approximated well by these limits.

\vspace{.1in} \noindent (d) The  pointer positions of the
apparatus  have the stability property as stated in section III
[item (iii)]. Different pointer positions have mutually disjoint
stability domains  in the phase space of the apparatus. When the
apparatus, after some interaction, is left to itself with its
`system point' (the point in the apparatus phase space
representing its instantaneous  classical state) in one of the
stability domains, it will eventually settle down to the
corresponding pointer position.

\vspace{.1in} \noindent (e) Different pointer positions
are macroscopically distinguishable [the
macroscopic distinguishability can be interpreted, for example, in
terms of an energy-time uncertainty product inequality ($\Delta E
\Delta t > > \hbar$) relevant to the experimental situation].

\vspace{.1in} \noindent (f) observations on the apparatus are
restricted to readings of the output devices (pointers).

 \vspace{.15in} \noindent  \textbf{\large 9. Concluding Remarks}

\vspace{.1in} \noindent 1. The central message of the present work
is this : `In the theoretical treatment of measurement on a quantum
system, the apparatus must be properly treated as a quantum
\emph{system}  approximated well by a classical one.' The main body
of the paper is devoted to just doing this job sensibly.

We have seen in operation the general plan (II, sections 4 and 6)
for dealing with situations involving quantum systems approximated
well by classical ones : Start with the quantum system (treating it
as a supmech Hamiltonian system), transform it to an equivalent
supmech Hamiltonian system, employing the Weyl-Wigner-Moyal
formalism and then introduce the classical approximations of the
relevant phase space functions. This made it possible (and smooth
going) the treatment of the apparatus along the above mentioned
lines.

This reduces the problem of `quantum-classical divide' in the
foundations of quantum mechanics to a non-problem --- there is no
need for such an ad hoc divide now.

\vspace{.1in} \noindent 2. It is worth (re-)emphasizing that, in the
theoretical treatment of quantum measurements, if the physics of the
apparatus is treated adequately (with due attentions to points
mentioned in section III), the  decoherence effects needed to wipe off
the unwanted quantum interferences appear automatically.  The sight
of Eq.(41) where one can see the operation of the decohering effect
of averaging over the passive region of the apparatus phase space
(which can be interpreted as the effect of the `internal
environment' of the apparatus) in live action, should please
theoreticians. The incorporation of the external environment (for
the restricted purpose of realizing von Neumann reduction) has been
reduced to a matter of less than two lines : just saying that the
symbol `A' now stands for `the apparatus and the external
environment'.

It is easily checked that the arguments against the traditional
decoherence program [7] do not apply in the present case. Here
the desired decoherence effects appear as a matter of course in a
consistent scheme of mechanics --- no questionable ad-hoc procedures
were employed to obtain those effects.

\vspace{.15in} \noindent \textbf{\large References}

\vspace{.12in} \noindent \begin{description}
\item[[1]] T. Dass,  A Stepwise Planned Approach to the Solution
of Hilbert's Sixth Problem. I : Noncommutative Symplectic Geometry
and Hamiltonian Mechanics. arXiv : 0909.4606  [math-ph] (2010).
\item[[2]] T. Dass,  A Stepwise Planned Approach to the Solution
of Hilbert's Sixth Problem. II : Supmech and Quantum Systems.
arXiv : 1002.2061  [math-ph] (2010).
\item[[3]] J. von Neumann,  \textsl{Mathematical Foundations of Quantum
Mechanics},  Princeton University Press, 1955.
\item[[4]] J.A. Wheeler, W.H. Zurek, \textsl{Quantum Theory and
Measurement}, Princeton University Press, 1983.
\item[[5]]  J.M. Jauch,  \textsl{Foundations of Quantum Mechanics}, Addison
Wesley, Reading, 1968.
\item[[6]] R. Omnes,  \textsl{The Interpretation of Quantum
Mechanics}, Princeton University Press, 1994.
\item[[7]] T. Dass,  Measurements and Decoherence, arXiv:
quant-ph/0505070 (2005).
\item[[8]] R. Omnes, \textsl{ Understanding Quantum Mechanics}, Princeton
University Press, 1999.
\item[[9]] W.H. Zurek, Decoherence, Einselection and Quantum Origin
of the Classical, \textsl{Rev. Mod. Phys.} \textbf{75}, 715-77 (2003).
\item[[10]] J. Bub, \textsl{Interpreting the Quantum World}, Cambridge
University Press, 1997.
\item[[11]] S.L. Adler, `Why decoherence has not solved the measurement
problem : a response to P.W. Anderson', arXiv : quant-ph/0306072 (2003).
\item[[12]] M. Dubois-Violette,  Noncommutative Differential
Geometry, Quantum Mechanics and Gauge Theory, in: \textsl{Lecture Notes in
Physics}, \textbf{375},  Springer, Berlin, 1991.
\item[[13]] M. Dubois-Violette,  Some Aspects of Noncommutative
Differential Geometry, in  \textsl{New Trends in Geometrical and Topological
Methods}, pp 145-157 (Madeira 1995);  arXiv: q-alg/9511027.
\item[[14]]  Y. Matsushima,  \textsl{Differentiable Manifolds}, Marcel Dekker,
New York, 1972.
\item[[15]] K. Gottfried,  \textsl{Quantum mechanics, volume I :
Fundamentals}, W.A. Benjamin, Inc., New York, 1966.
\item[[16]] P. Busch, M. Grabowski, P.J. Lahti,
\textsl{Operational Quantum Physics}, Springer-Verlag, Berlin, 1995.
\item[[17]] C. Cohen-Tannoudji, B. Diu, F. Lalo$\ddot{e}$,
\textsl{Quantum Mechanics, vol I}, John Wiley and Sons, 2005.
\item[[18]] G. Potel, et al,  Quantum Mechanical Description of Stern-Gerlach
Experiments. arXiv : quant-ph/0409206 (2004).
\item[[19]] A. Goswami,  \textsl{Quantum mechanics}, Wm. C. Brown Publishers, 1992.
\end{description}

\end{document}